\newdimen\rotdimen
\def\vspec#1{\special{ps:#1}}
\def\rotstart#1{\vspec{gsave currentpoint currentpoint translate
   #1 neg exch neg exch translate}}
\def\rotfinish{\vspec{currentpoint grestore moveto}}
\def\rotr#1{\rotdimen=\ht#1\advance\rotdimen by\dp#1%
   \hbox to\rotdimen{\hskip\ht#1\vbox to\wd#1{\rotstart{90 rotate}%
   \box#1\vss}\hss}\rotfinish}
\def\rotl#1{\rotdimen=\ht#1\advance\rotdimen by\dp#1%
   \hbox to\rotdimen{\vbox to\wd#1{\vskip\wd#1\rotstart{270 rotate}%
   \box#1\vss}\hss}\rotfinish}%
\def\rotu#1{\rotdimen=\ht#1\advance\rotdimen by\dp#1%
   \hbox to\wd#1{\hskip\wd#1\vbox to\rotdimen{\vskip\rotdimen
   \rotstart{-1 dup scale}\box#1\vss}\hss}\rotfinish}%
\def\rotf#1{\hbox to\wd#1{\hskip\wd#1\rotstart{-1 1 scale}%
   \box#1\hss}\rotfinish}%
\documentstyle[12pt,world_sci,epsfig]{article}
\begin{document}

\vspace*{-1.7cm}
\rightline{\vbox{\halign{&#\hfil\cr
& OCIP/C-93-12\cr
& September 1993\cr}}}

\title{{\bf HEAVY CHARGED LEPTON PAIR PRODUCTION\\
THROUGH PHOTON FUSION AT HADRON SUPERCOLLIDERS}
\thanks{To be published in
the Proceedings of the {\sl Madison-Argonne Workshop on Physics at Current
Accelerators and the Supercollider}, March - June 1993}
\thanks{Also available via anonymous ftp from ftp.physics.carleton.ca
as /pub/theory/peterson/ocipc9312.ps}
}
\author{G.\ BHATTACHARYA, P.\ A.\ KALYNIAK, and K.\ A.\ PETERSON\\
{\em Ottawa-Carleton Institute for Physics\\
Department of Physics, Carleton University,
1125 Colonel By Drive\\
Ottawa, Ontario, CANADA K1S 5B6}\\}

\maketitle

\begin{center}
\parbox{13.0cm}
{\begin{center} ABSTRACT  \end{center}
{\small \hspace*{0.3cm} The pair production of charged heavy leptons via two
photon fusion is considered for hadron collisions at SSC and LHC energies.
Rates for the inelastic process $pp \rightarrow \gamma \gamma X \rightarrow
L^+L^-X$ and the elastic process $pp \rightarrow \gamma \gamma pp
\rightarrow L^+L^- pp$ are given in a Weizs\"{a}cker-Williams
approximation and compared with production via the gluon fusion and Drell-Yan
mechanisms}}
\end{center}

\section{Introduction}

Heavy leptons, both charged and neutral, are a feature of many models which
extend the particle
content of the Standard Model. These include models which propose a
complete additional
generation\cite{1} of heavy quarks and leptons, as well as those like $E_6$
based grand unified
models\cite{2}, which contain extra particles
within each generation. The
Drell-Yan
\cite{3,4}, gluon fusion\cite{4}, and gauge boson fusion\cite{5} mechanisms
for the production
of heavy charged leptons in hadron collisions have been investigated in the
past.

We present here
a preliminary study of heavy charged lepton pair production through two
photon fusion, which has been
overlooked so far.
Both the inelastic process $pp \rightarrow \gamma \gamma X \rightarrow L^+
L^-X$ and the
elastic process, $p p \rightarrow \gamma \gamma p p \rightarrow L^+L^- p p$,
are considered here
in a Weizs\"{a}cker-Williams approximation (WWA)\cite{6,7}. Our investigation
indicates that
two photon fusion is an important mechanism for the production of charged
lepton pairs, at the SSC, provided the lepton mass, $m_L$, is below about 250
GeV.

For comparison with the photon fusion mechanism considered here, we would also
present the results for $L^+L^-$ production through the Drell-Yan and
gluon fusion mechanisms.
The Drell-Yan quark anti-quark annihilation proceeds
via s-channel $\gamma$ and $Z$ exchange. The gluon fusion proceeds via a quark
triangle diagram followed by $Z$ or
Higgs exchange. Expressions for these cross sections can be found in the
literature \cite{4,8}. The rates for the gauge boson fusion and annihilation
processes are negligible compared to the Drell-Yan
and gluon fusion rates except for the case of very massive heavy
leptons\cite{5}
. Since we find the two photon fusion mechanism to be important only for
relatively small values of $m_L$, further comparisons with the gauge boson
fusion processes will not be made.

For the two photon fusion mechanism under consideration here, the Feynman
diagrams for the relevant subprocess $\gamma\gamma \rightarrow L^+L^-$, common
to the inelastic and elastic cases,
 are
shown in Fig. 1. Assuming that heavy charged leptons couple to the photon in
the usual way, the summed and averaged matrix element squared for this
process is

\begin{eqnarray}
\sum \overline{|{\cal M}|^2} & = & 2 e^4 \biggl[\frac{(u-m_L^2)}{(t-m_L^2)}
+ \frac{(t-m_L^2)}{(u-m_L^2)}
- 4 m_L^2 \Bigl(\frac{1}{(t-m_L^2)}+\frac{1}{(u-m_L^2)}\Bigr) \nonumber \\
& &- 4 m_L^4 \Bigl(\frac{1}{(t-m_L^2)} +\frac{1}{(u-m_L^2)}\Bigr)^2\biggr]
\end{eqnarray}

\noindent where $t$ and $u$ refer to the exchanged momenta squared
corresponding to
the direct and crossed diagrams for the two photon subprocess. The total
cross sections for the elastic and inelastic cases are obtained by convoluting
the subprocess cross section
with the appropriate photon and parton structure functions and integrating
over the two body phase space using Monte Carlo techniques.

\begin{center}
\mbox{ \epsfig{figure=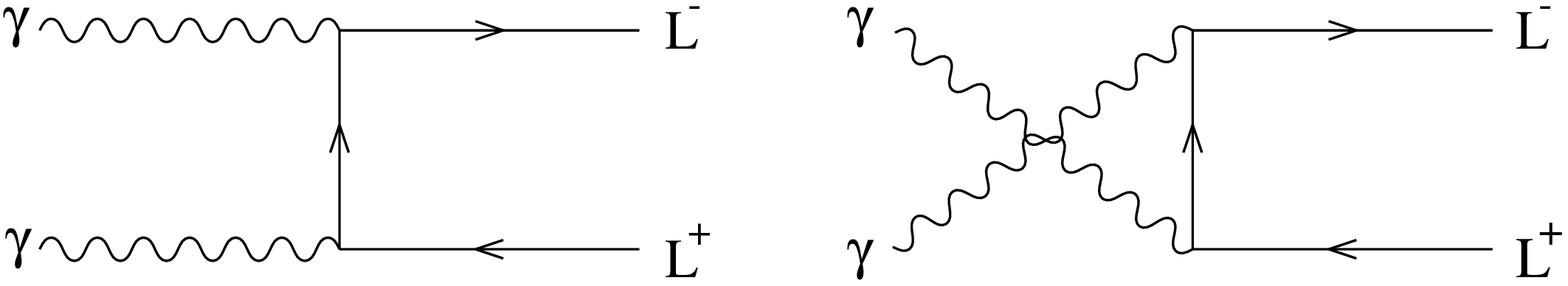,height=10cm}}
\end{center}
{\small Fig. 1. The Feynman diagrams which contribute to the subprocess
$\gamma \gamma \rightarrow L^+L^-$.}
\vspace{0.5cm}

We describe in Secs. 2 and
3 respectively, the details of the total cross section calculations
for the inelastic and elastic cases. The results are presented in Sec. 4
and we summarize in the final section and indicate
what work is in progress.

\section{The Inelastic Process $p p \rightarrow \gamma \gamma X \rightarrow L^+
L^- X$}

The cross section for the production of a pair of heavy charged
leptons in inelastic $pp$ collisions is obtained by
convoluting the differential cross section of the $\gamma \gamma \rightarrow
L^+ L^-$ subprocess, $d\hat{\sigma}_{\gamma \gamma}$, with the
probabilities of finding the
photons in the protons as follows.

\begin{equation}
d \sigma^{inel} (s) = \int_{x_{\rm min}}^{1} dx_1
\int_{x_{\rm min}/x_1}^{1} dx_2 \, f^{inel}_{\gamma / p} (x_1) \,
f^{inel}_{\gamma / p} (x_2) \, d\hat{\sigma}_{\gamma \gamma}(x_1 x_2 s)
\end{equation}

The inelastic component of the photon spectrum, $f^{inel}_{\gamma / p} (x)$,
is given by

\begin{equation}
f^{inel}_{\gamma / p} (x) = \int_{x_{\rm min}}^{1} dx_1
\int_{x_{\rm min}/x_1}^{1} dx_2 \, \sum_{q}
f_{\gamma / q} (x_1) \, f_{q / p} (x_2, Q^2) \, \delta (x - x_1 x_2)
\end{equation}

\noindent
In the above equation, $f_{q / p} (x_2, Q^2)$ are the parton
structure functions, which we have taken to be the HMRS (Set B) structure
functions\cite{9} evaluated at the scale $Q^2 = \hat{s}/4$, $\hat{s}$ being
the parton center of mass energy squared. We sum over the contributions due to
the $u, d, c$, and $s$ quarks and antiquarks from the protons.
The lower limits of integration in the above equations ensures that
the two photon center of
mass energy is sufficient for $L^+L^-$ production. The Weizs\"{a}cker-Williams
photon spectrum, $f_{\gamma / q}(x)$, from a quark (of charge $e_q$) is given
by

\begin{equation}
f_{\gamma / q} (x) = \frac{e_{q}^2}{8 \pi^2} \frac{1 + (1 - x)^2}{x} \log
\frac{t_{max}}{t_{min}}.
\end{equation}

\noindent
Here $t_{max}$ and $t_{min}$ are the characteristic maximum and minimum photon
momentum
transfers which
we have taken to be
$t_{max} = \hat{s} /4 - m_{L}^2$ and $t_{min} = 1$ ${\rm GeV}^2$, with
$\sqrt{\hat{s}}$ being
the center of mass energy in the parton frame. There is some flexibility in
the choice of $t_{max}$. However, in agreement with
Altarelli et. al. \cite{10}, we have found that our results are not very
sensitive to this
parameter, within the limits of the Weizs\"{a}cker-Williams approximation.
The particular choice of the minimum momentum transfer, $t_{min}$,
guarantees that the photons are obtained from the deep inelastic scattering
of protons, when the quark-parton model is valid.

\section{The Elastic Process $p p \rightarrow \gamma \gamma p p \rightarrow L^+
L^- p p$}

The cross section for $L^+L^-$ production via elastic collisions of protons
is obtained by folding the same subprocess differential cross section, $d\hat
{\sigma}_{\gamma \gamma}$, with
the elastic component of the photon spectrum from the protons, $f^{el}_{\gamma
/ p}$.

\begin{equation}
d \sigma^{el} (s) = \int_{x_{\rm min}}^{x_{\rm max}} dx_1 \int_{x_{\rm
min}/x_1}^{x_{\rm max}} dx_2 \, f^{el}_{\gamma / p}
(x_1) \,
f^{el}_{\gamma / p} (x_2) \, d\hat{\sigma}_{\gamma \gamma}(x_1 x_2 s)
\end{equation}

\noindent The upper limits of integration in the above expression are given by
$x_{\rm max} = 1 - 2 m_p/{\sqrt s}$, where $m_p$ and $\sqrt s$ are respectively
the proton mass and the center of mass energy of the elastically colliding
protons. The lower limit of integration $x_{\rm min}$ has the same significance
as in the case of inelastic collisions. In the above, we use the photon
spectrum for elastic collisions of protons, $f^{el}_{\gamma / p}$, derived by
Kniehl\cite{7} in a modified Weizs\"{a}cker-Williams approximation. It takes
the form given below.

\begin{equation}
f^{el}_{\gamma / p} (x) = -\frac{\alpha}{2 \pi} x \int_{-\infty}^{t_{max}}
\frac{dt}{t}
\left \{2 \left[ \frac{1}{x} \left( \frac{1}{x} - 1 \right) + \frac{m_p^2}{t}
\right]
H_1(t) + H_2(t) \right \}
\end{equation}

\noindent Here, $t_{max} = -m_p^2 x^2/(1 - x)$ while $H_1$ and $H_2$  are
functions of the
electric and magnetic form factors of the proton.

\begin{equation}
H_1(t) = \frac{G_E^2(t) -(t/4m_p^2)G_M^2(t)}{1 - t/4m_p^2} , \; \;\; \; H_2(t)
= G_M^2(t)
\end{equation}

\noindent The form factors are parametrized as

\begin{equation}
G_E(t) = (1 - t/0.71 \, {\rm GeV}^2)^{-2} , \; \;\; \; G_M(t) = 2.79 G_E(t).
\end{equation}

{}From the above one can obtain an explicit expression for $f^{el}_{\gamma /p}$
which has been used in our calculation. The full details are given in Ref. 7.
We have also checked the above calculation by using an alternative form of the
photon spectrum from elastically colliding protons \cite{11}. The cross
sections thus obtained using this alternative form essentially agree with the
results presented here using the Kniehl form of WWA.

\section{Results and Discussion}
For the purpose of comparison, we first present some results for the
two previously well known production
mechanisms for $L^+L^-$, Drell-Yan and gluon fusion. In our calculations
we use $\alpha = 1/128$, $M_W = 80.22$ GeV, and $M_Z = 91.173 \,$ GeV.
The total cross sections are obtained by convoluting the parton subprocess
cross sections with the HMRS (Set B) structure functions evaluated at the
scale $Q^2 = \hat{s}/4$, where $\sqrt {\hat s}$ is the center of mass energy
for the subprocess. The parameter $\alpha_s$, used in the gluon fusion
calculation, is evaluated at the two-loop
level using its representation in the modified minimal-subtraction $(\overline
{\rm MS})$ scheme\cite{12}, and using the value $\Lambda_{\overline {\rm MS}}^
{(4)} =
0.19$ GeV, consistent with the HMRS-B parametrization.
For the gluon fusion mechanism we assume only three
generations of quarks, with the top quark mass set at 150 GeV. Results for
the inclusion of a fourth generation of heavy quarks can be found in Refs. 1,8.
The Higgs mass for the calculation is chosen to be 150 GeV.
\vbox{
\begin{center}
\vspace{-3.5cm}
\setbox1=\vbox{\mbox{\epsfig{figure=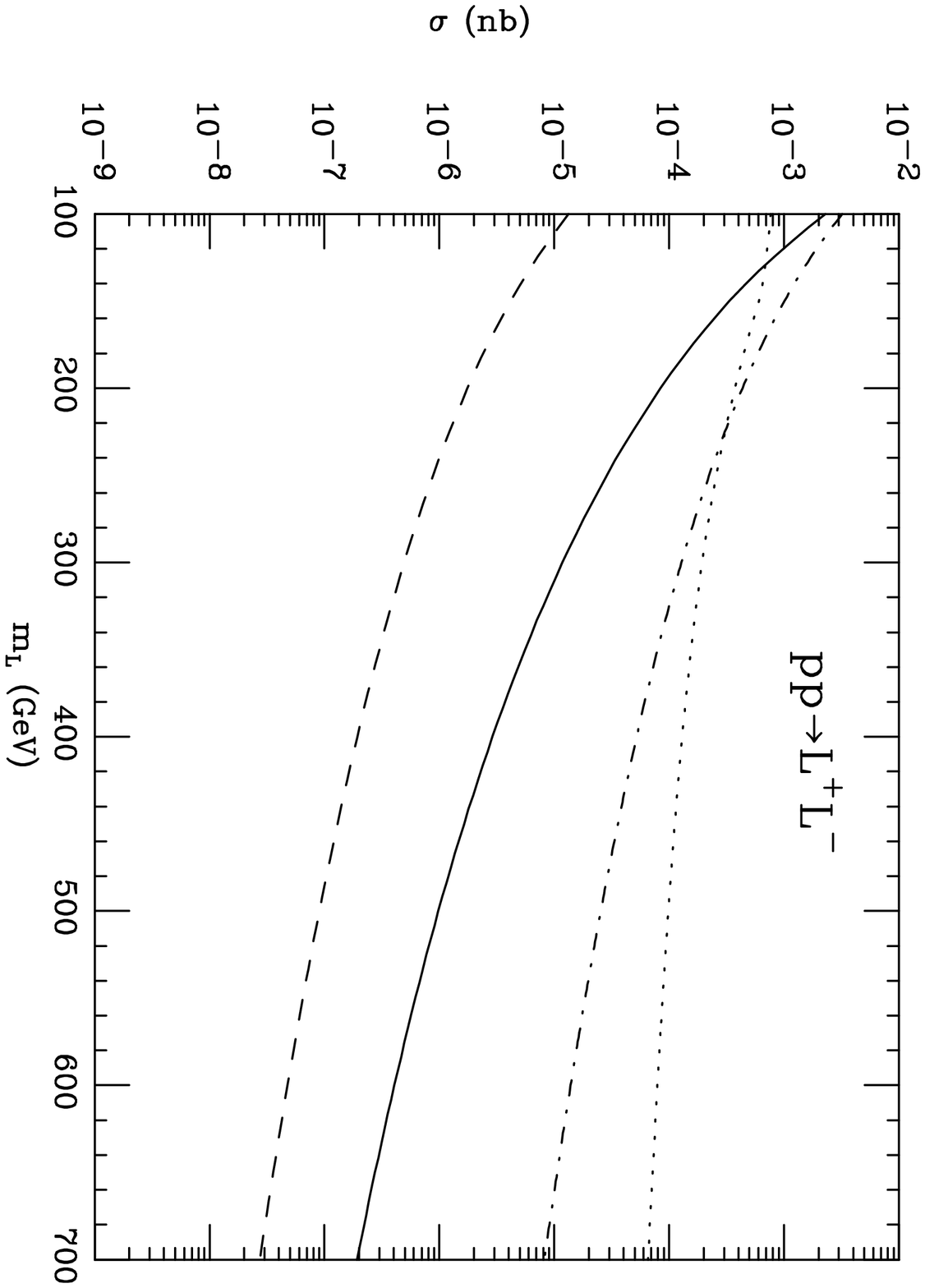,height=10cm}}}
\rotl{1}
\vspace{-4.0cm}
\end{center}
\noindent
{\small Fig. 2 The total production cross section (in nanobarns) for a
charged lepton pair in $pp$ collisions at the SSC ($\sqrt s = 40$ TeV) as a
function of its mass $m_L$. The solid curve represents the lepton pair
production through two photon fusion in the deep inelastic scattering region
of protons, and the dashed curve shows the photon fusion production of $L^+
L^-$ for elastic collision of protons. The dotted and dot-dashed
curves represent respectively production through gluon fusion and the Drell-Yan
mechanisms.}
}

The total cross sections for $L^+L^-$ production via the Drell-Yan and gluon
fusion processes in $pp$ collisions, as a function of the charged lepton
mass, are shown in Fig. 2 for the SSC center of mass energy of 40 TeV and in
Fig. 3 for the LHC energy of 15.4 TeV. The dotted and dot-dashed lines
represent respectively the cross sections due to gluon fusion and Drell-Yan
mechanisms.
 At both energies, the Drell-Yan process
dominates for low
$m_L$. However, the gluon fusion cross section falls off less rapidly than the
Drell-Yan cross section with
increasing $m_L$, and
overtakes the Drell-Yan around $m_L =$ 240 GeV for the SSC, and at about
$m_L =$ 500 GeV for the
LHC.

In Figs. 2 and 3, the solid lines represent the total cross section for the
production of charged heavy
leptons via the inelastic process $p p \rightarrow \gamma \gamma X \rightarrow
L^+ L^- X$, as a
function of $m_L$, for the SSC and LHC energies respectively. Evidently, the
inelastic two photon
process is an important means of producing fairly low mass heavy charged lepton
pairs at SSC
energies. Fig. 2 shows that the cross section for this process is within a
factor of 1.4 of that for
the dominant Drell-Yan process when $m_L$ is 100 GeV. The two photon
inelastic production
falls to an order of magnitude below the now-dominant gluon fusion process by
$m_L$ of about 260
GeV. Hence, we can conclude that the process $p p \rightarrow \gamma \gamma X
\rightarrow L^+ L^- X$ is an important means of production of heavy charged
lepton pairs for
$m_L$ below 200-250 GeV at SSC energies. From Fig. 3, we can see that the
inelastic two photon
process under consideration is much less important at LHC energies. The
production cross
section is over an order of magnitude below the dominant Drell-Yan production
cross section even for $m_L$ of 100
GeV.
\vbox{
\begin{center}
\vspace{-3.5cm}
\setbox1=\vbox{\mbox{\epsfig{figure=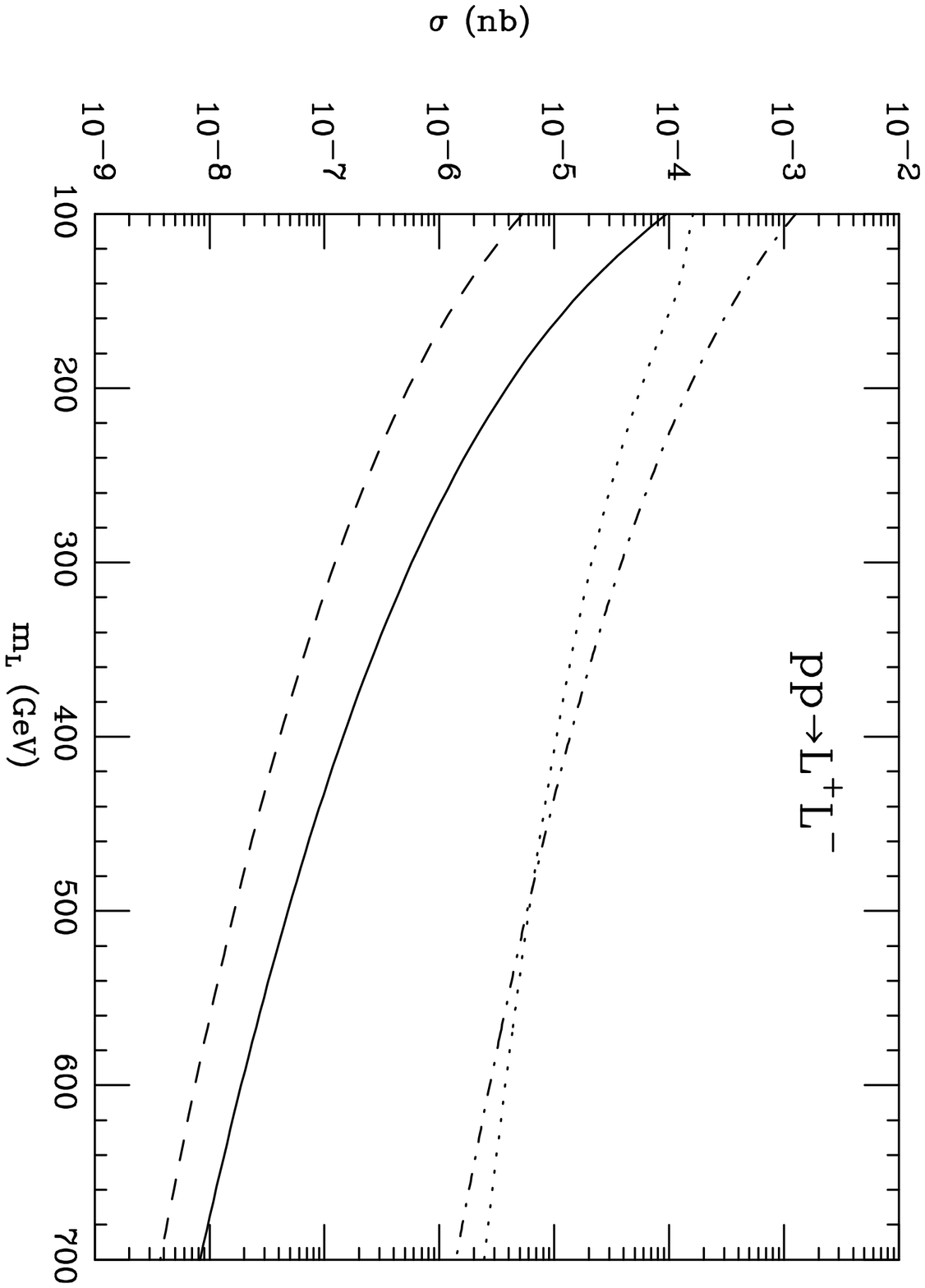,height=10cm}}}
\rotl{1}
\vspace{-4.0cm}
\end{center}
{\small Fig. 3 The total production cross section (in nanobarns) for a charged
lepton pair in $pp$ collisions at the LHC ($\sqrt s = 15.4$ TeV) as a function
of its mass $m_L$. The different curves represent the same processes as in
Fig. 2}
}
\newpage
\noindent
The total cross sections for the elastic $pp$ collision is given as a function
of the mass of the heavy
lepton, by the dashed lines in Figs. 2 and 3, for the SSC and LHC energies
respectively. This
contribution is well below the dominant contributions over the range of masses
under consideration at both energies.
 The inelastic cross section is more than an order of magnitude larger
than the
elastic cross section for relatively low $m_L$ at SSC energies, but their
difference
narrows with increasing $m_L$, and also as
the energy is reduced to
that of the LHC.
We note that the inelastic and elastic contributions are comparable at
Tevatron energies. However, for the
high energy
colliders which we consider here, the elastic two photon production of charged
heavy lepton
pairs is not very important.

\section{Conclusions}

We have found that the inelastic two photon fusion production of pairs of heavy
charged leptons
in hadron colliders, $p p \rightarrow \gamma \gamma X
 \rightarrow L^+ L^- X$, can be an important mechanism relative to the
Drell-Yan and gluon
fusion production mechanisms. Our calculation is done in a
Weizs\"{a}cker-Williams approximation.
At SSC energies, the inelastic two photon fusion is found to be comparable to
the dominant
Drell-Yan mechanism for a heavy lepton mass of 100 GeV. The two photon fusion
cross section
drops more rapidly than that of the Drell-Yan process as the heavy lepton mass
increases. Hence,
at SSC energies, the photon fusion mechanism is important provided the lepton
mass is below
about 200 - 250  GeV. At LHC energies, the inelastic two photon fusion cross
section is at least
an order of magnitude below the dominant process cross sections over the mass
range which we
consider. Further, we have also calculated the elastic two photon fusion
process, $p p
\rightarrow  \gamma \gamma p p \rightarrow L^+L^- p p$, within a modified
Weizs\"{a}cker-Williams  approximation, and find it to contribute at most 10\%
as much as the
inelastic process at SSC energies. At LHC energies, it is also negligible
relative to the
dominant processes. However the elastic process might offer a very clean signal
for detection of charged heavy leptons, as opposed to the other processes
discussed here.

We are in the process of doing the exact calculation for both the inelastic
and elastic
processes considered here in order to verify the validity of our
Weizs\"{a}cker-Williams approximations in the two cases.

\begin{center}
ACKNOWLEDGEMENTS
\end{center}

\noindent
This work was supported in part by the Natural Sciences and Engineering
Research Council of Canada. One of us (P.A.K.) thanks the Deans of Science and
Graduate Studies and Research at Carleton University for a travel grant. We
thank A. Soni for suggesting that the two photon fusion mechanism might be
important, and D. Zeppenfeld for helpful discussions on Effective Photon
Approximation for elastic collision of protons. One of us (G.B.)
acknowledges the warm hospitality of the Institute for Elementary Particle
Physics Research at the University of Wisconsin-Madison during the later stages
of completion of this work.

\bibliographystyle{unsrt}

\end{document}